\documentclass{article}

\usepackage[utf8]{inputenc}
\usepackage{graphicx}
\graphicspath{{Images/}}
\usepackage{color, soul}
\usepackage{amsmath, amsfonts, amssymb, amsthm}
\usepackage{dsfont}
\usepackage{stmaryrd}
\usepackage{url}
\usepackage{array}

\usepackage{nomencl}
\makenomenclature

\newtheorem{theorem}{Theorem}
\newtheorem{property}{Property}
\newtheorem{definition}{Definition}
\newtheorem{remark}{Remark}
\newtheorem*{example}{Example}

\title{Graph model selection by edge probability sequential inference}
\author{Louis Duvivier \and Rémy Cazabet \and Céline Robardet}

\begin{document}

\maketitle

\section*{Abstract}
Graphs are widely used for describing systems made up of many interacting components and for understanding the structure of their interactions. Various statistical models exist, which describe this structure as the result of a combination of constraints and randomness.
In this paper, we introduce edge probability sequential inference, a new approach to perform model selection, which relies on probability distributions on edge ensembles. From a theoretical point of view, we show that this methodology provides a more consistent ground for statistical inference with respect to existing techniques, due to the fact that it relies on multiple realizations of the random variable. It also provides better guarantees against overfitting, by making it possible to lower the number of parameters of the model below the number of observations. Experimentally, we illustrate the benefits of this methodology in two situations: to infer the partition of a stochastic blockmodel, and to identify the most relevant model for a given graph between the stochastic blockmodel and the configuration model.

\tableofcontents


\section{Introduction}
Graphs are a powerful mathematical abstraction to study interconnected objects. Beyond the nature of interactions between individuals, which range from chemical reactions between substrates in a metabolic network \cite{holme2003subnetwork} to co-authorship between scientists \cite{newman2004coauthorship}, to neurons connections in our brain \cite{rubinov2010complex} and many others, their overall structure provides information on the system. In particular, graphs coming from various domains have been shown to exhibit similar topological properties, such as short paths between nodes, a long-tailed degree distribution, high clustering coefficient or a modular structure \cite{costa2007characterization}.

Statistical models aim to explain these structures as the result of a random sampling of the graph subject to some constraints \cite{park2004statistical}. It does so by defining a set of graphs and a probability distribution such that a given property is verified, by all graphs in the set (microcanonical ensemble) or on average (canonical ensemble), as detailed in \cite{cimini2019statistical}. This allows to distinguish whether the structure of the graph is a mere consequence of the imposed constraints or if additional explanations are required.

Various models have been proposed, which rely on different properties of real world graphs obtained from observations. The simplest one, which is also the oldest, is the Erdös-Rényi model, which imposes as only constraints the number of nodes and edges in the graph \cite{erdos1960evolution}. The observation of graphs, whose degree distribution differs significantly from what would be expected according to Erdös-Rényi model, has led to the study of random graphs with imposed degree distribution \cite{newman2001random}, known as the configuration model. Among the more widely used models is also the stochastic blockmodel \cite{soderberg2002general}, which partitions nodes into groups and constrains the number of edges between each of these groups. An overview of other existing statistical graph models can be found in \cite{goldenberg2010survey}. 

This variety of parametric probabilistic models raises the issue of model selection: given a graph $G$, which model best describes its edge distribution, and with which set of parameters? This question has particularly been studied in the domain of community detection, which can be understood as: assuming that the connections between nodes are determined by an underlying node partition, what is this partition? Various objective functions have been proposed for this task, the most famous one been the modularity \cite{newman2004finding}, but it has been shown that they may identify communities even in random graphs, thus mistaking noise for structure \cite{guimera2004modularity}. 

Statistical models provide a way to overcome this lack of theoretical guarantees. By defining models as probability distributions on graph ensembles, they offer a natural measure of their complexity, the entropy, which makes them comparable. Indeed, some work has been devoted to the computation of the entropy of graph ensembles. In \cite{bianconi2007entropy}, the author computes the entropy of canonical ensembles of directed and undirected random graphs. In \cite{bianconi2009entropy} she extends these results to the case of weighted graphs and prescribed degree distribution. In \cite{peixoto2012entropy}, the author computes the entropy of both microcanonical and canonical stochastic blockmodel ensembles, with and without degree correction. In \cite{coon2018entropy}, authors consider the case of spatial graph ensembles, and in \cite{zingg2019entropy}, they tackle generalized hypergeometric random graph ensembles.

What is more, this probabilistic definition allows to rigorously compute the most likely set of parameters used to generate a given observation, thanks to bayesian inference. This has been applied to community detection in \cite{prokhorenkova2019community}, by defining a stochastic blockmodel with a given set of parameters as a probability distribution on a set of graphs, and applying Bayes' theorem to compute the most likely set of parameters given an observed graph. In \cite{peixoto2019bayesian}, the author leverages the fact that in the microcanonical ensemble, the maximisation of the likelihood of a set of parameters is equivalent to the minimization of the entropy of its associated probability distribution to perform inference.

However, these works relies on probability distributions defined on sets of graphs, which means that the observation of a single graph (which in practice is the most common situation) corresponds to a single realization of the random variable. Even though bayesian inference requires less observations than frequentist inference to be sound, a single realization induces a high risk of overfitting. It is also not trivial to adapt this methodology to compare not only different sets of parameters (such as node partitions in the stochastic blockmodel), but models of a different nature (such as a stochastic blockmodel and a configuration model). 

In this paper, we introduce an alternative point of view on graph statistical models, which relies on probability distributions defined on sets of edges. Because a single graph contains many edges, it implies that the same observed graph corresponds to many realizations of the random variable rather than just a single one. As a consequence, the inference of the underlying probability distribution is more rigorous. Indeed, it allows to control the number of parameters of the model in order to ensure that it remains below the number of observations, which is a necessary condition to avoid overfitting. Moreover, as it formulates all statistical models in terms of probability distributions on the same set of edges, it provides a natural framework to compare them and find the most relevant one with respect to a given graph.

The paper is organized as follows. In section~\ref{edge statistical model}, we introduce edge statistical models and explain how they differ from usual graph statistical models. In section~\ref{model_selection}, we develop sequential edge probability inference, a theoretical framework to perform inference using probability distribution on sets of edges. We then illustrate in section~\ref{application to model selection} how it can be used both to infer the parameters of a statistical model (subsection \ref{sbm partition}) and to compare stochastic blockmodel and configuration model with respect to a given graph (subsection \ref{sbm and cfm}). Section~\ref{sec:conclusion} concludes and gives directions for future work.

\section{Edge statistical model}
\label{edge statistical model}
\subsection{Definition}

Statistical models aim at describing the distribution of edges in a graph as the result of a random process subject to some constraints. This can be done in various manner. Classical statistical graph models are usually defined as a set of graphs $\Omega_M$ and a probability distribution $\mathbb{P}_M$ on this set. There exists two main ways to define such models, inspired from statistical physics: microcanonical and canonical ensembles \cite{cimini2019statistical}. Both rely on a property $P$, which depends on the model. Some examples of graph models and there associated properties are given in Table \ref{graph statistical models}. 

\begin{table}[htb]
    \centering
    \begin{tabular}{l|l}
         Model & Property \\
         \hline
         Erdos Reyni & number of edges $m$ \\
         \hline
         Configuration Model & degree distribution $(k_i)_{i \in [1,n]}$ \\
         \hline
         Stochastic blockmodel & node partition $B = (b_1, \dots, b_p)$ \\
          & block adjacency matrix $W \in M_p(\mathbb{N})$ \\
          \hline
    \end{tabular}
    \caption{Examples of graph statistical model.}
    \label{graph statistical models}
\end{table}
For a property $P$, and a given value of this property $p_0$ (which can be learned on an observed graph) the microcanonical model is of the form
\[\Omega_{M(P)} = \{G = (V,E) \mid P(G) = p_0\}\]
and $\mathbb{P}_{M(P)}$ is defined as the uniform distribution on this set. On the other hand, the canonical model is of the form 
\[\Omega_{M(P)} = \{G = (V,E)\} \] 
and $\mathbb{P}_{M(P)}$ is the maximum entropy distribution such that $\mathbb{E}[P(G)] = p_0$. In both cases, the random variable, whose probability distribution is studied, is a graph. Thus, we call this type of models \textbf{graph statistical models}.
As in practice we almost always study a single graph, the problem with such a statistical model definition is that statistical inference involves to fit a probability distribution on a single realization of the random variable, which implies a high risk of overfitting.

To overcome this issue, we assume that edges are generated independently from each other. Therefore, a model $M$ can be defined as a probability distribution $\mathbb{P}_M$ on the set of possible edges $\llbracket 1,n \rrbracket^2$ (For the simplicity of computations, we consider directed graphs and authorize self-loops but the methodology could easily be adapted for undirected edges and forbidden self-loops by restricting this set of possible edges). We will call this type of models \textbf{edge statistical models}, because the probability distribution is defined on a set of edges. As edges are assumed to be independent, an ordered sequence of edges $E = (e_1, \dots, e_m)$ is generated with probability:
\[ \mathbb{P}_M[E] = \prod_{i = 1}^m \mathbb{P}_M[e_i] \]%
\nomenclature{$n, m$}{number of nodes, edges}{}{}
\nomenclature{$E = (e_1, \dots, e_m)$}{sequence of edges}{}{}

\begin{center}\fbox{\begin{minipage}{0.95\textwidth}
\begin{example}
Let's take some examples to illustrate how frequently used statistical models can be formulated as probability distribution on edges. The simplest model is the fully random Erdos-Reyni. It corresponds to the uniform distribution on $\llbracket 1,n \rrbracket^2$:
\[ \forall u,v \in [1,n], \mathbb{P}_{ER(n)}[u,v] = \frac{1}{n^2}\]
Then, the configuration model: instead of a degree sequence, it takes as parameter a probability distribution $(p_i)_{i \in [1,n]}$ corresponding for each node to its probability of being picked at random as an extremity of the generated edge:
\[ \forall u,v \in [1,n], \mathbb{P}_{CFM((p_i)_i)}[u,v] = p_u \times p_v\]
It's directed version is straightforward, considering the probability distributions $(p_i^{out})_i$ and $(p_i^{in})_i$. 

Finally, the stochastic blockmodel takes as parameter a partition $B = (b_1, \dots, b_p)$ and a block probability matrix $P \in M_p([0;1])$ such that $\sum_{i,j} |b_i||b_j|P_{i,j} = 1$. If $u \in b_i$ and $v \in b_j$, the edge $(u,v)$ is generated with probability:
\[ \mathbb{P}_{SBM(B,P)}[u,v] = P_{i,j} \]
\label{sbm_def}
\end{example}
\end{minipage}}
\end{center}

Edge statistical models naturally generate temporal multigraphs, in which edges are ordered and each edge may appear multiple times. Indeed, even if a given edge $(u,v)$ has already been sampled, its probability to be sampled again is not null. This is a natural way to model many real life interactions, even though this type of graphs is not the most widely used in practice.
Fortunately, edge statistical model adapts easily for static and simple graphs, since a static graph can be considered as the trace of a temporal one, in which edge ordering has been dropped.

\begin{definition}
    We say that an edge sequence $E = (e_1, \dots, e_m)$ \emph{collapses} to a static multigraph $G$, described by its weight matrix $W_G$ iff:
    \[ \forall u,v \in \llbracket 1, n \rrbracket, W_G[u,v] = |\left\{k \in [1, m]\mid e_k = (u,v)\right\}|\]
    We denote this $E \downarrow G$, and for any static multigraph $G$ we define the set of edge sequences which collapse to it by
    \[\mathcal{E}^{\downarrow}_G = \{E \mid E \downarrow G\}\]
\end{definition}

When studying a static multigraph $G$, one does not know from which sequence $E \in \mathcal{E}^{\downarrow}_G$ it derives. Studying all sequences would be very demanding as the size of the set is the multinomial coefficient $\binom{m}{w_1, \dots, w_{n^2}}$. However, as all edge sequences in $\mathcal{E}^{\downarrow}_G$ contain the same edges with the same multiplicity, by definition we have:
\begin{align*}
    \forall E_0 \in \mathcal{E}^{\downarrow}_G, \: \mathbb{P}_M[G] &= \sum_{E \in \mathcal{E}^{\downarrow}_G} \mathbb{P}_M[E] \\
    &= |\mathcal{E}^{\downarrow}_G| \times \mathbb{P}_M[E_0]
\end{align*}
This means that any sequence $E_0 \in \mathcal{E}^{\downarrow}_G$ can equivalently be chosen as a representative of $G$.

Beyond edge ordering, considering a simple graph means that we also discard edge multiplicity. 

\begin{definition}
    We say that an edge sequence \emph{simplifies} to a static simple graph $G$ described by its adjacency matrix $A_G$ iff:
    \[ \forall u,v \in \llbracket 1, n \rrbracket, A_G[u,v] = \mathds{1}_{(u,v) \in E}\]
    We denote this $E \Downarrow G$, and for any static simple graph $G$ with $m$ edges, we define the set of edge sequences which simplify to it, by:
    \begin{align*}
        \mathcal{E}^{\Downarrow k}_{G} &= \{|E| = m+k \mid E \Downarrow G\} \\
        \mathcal{E}^{\Downarrow}_{G} &= \bigcup_{k \geq 0} \mathcal{E}^{\Downarrow k}_{G}
    \end{align*}
\end{definition}

The number of edge sequences of length $(m+k)$ which simplify to $G$ grows exponentially with $k$ as $m!m^k \leq |\mathcal{E}^{\Downarrow k}_{G}| \leq m^{m+k}$. On the other hand, the probability to sample longer sequences decreases exponentially with $k$
\[ \forall M, \forall E \in \mathcal{E}^{\Downarrow k}_{G}, \mathbb{P}_M[E] \leq \prod_{e \in G} \mathbb{P}_M[e] \times p_0^k \text{ with } p_0 = \max_{e \in G} \, \mathbb{P}_M[e]\]
Therefore, as long as we consider models $M$ such that $\exists K, \max_{e \in G} \, \mathbb{P}_M[e] \leq \frac{K}{n^2}$ and $\frac{m}{n^2} \ll \frac{1}{K}$, the weight of $\mathcal{E}^{\Downarrow k}_{G}$ decreases exponentially with $k$ in $\mathcal{E}^{\Downarrow}_{G}$. Thus, we assume that the weight is concentrated on $\mathcal{E}^{\downarrow}_G = \mathcal{E}^{\Downarrow 0}_{G}$ and that we can choose a representative of $G$, $E_0 \in \mathcal{E}^{\downarrow}_G$.

\subsection{Edge probability distribution statistical inference}
\label{statistical_inference}

As an edge statistical model is defined as a probability distribution on $\llbracket 1,n \rrbracket^2$, an edge sequence $E$ corresponds to $m$ independent realizations of a random variable following the same unknown probability distribution $\mathbb{P}_0$. The objective of statistical inference is to make an estimation $\mathbb{Q}^*(E)$ of $\mathbb{P}_0$, avoiding both overfitting and underfitting, among the set of all possible models.

\begin{definition}
Let $\mathcal{M}_n^{\bullet}([0,1])$ be the set of all probability distributions on $\llbracket 1,n \rrbracket^2$:
\[ \mathcal{M}_n^{\bullet}([0,1]) = \left\{ \mathbb{Q} \in \mathcal{M}_n([0,1]) \mid \sum_{u,v \in \llbracket 1,n \rrbracket^2} \mathbb{Q}[u,v] = 1 \right\}.\]
Its elements can be seen as $n \times n$ matrices or as probability distributions. In the following we will use both points of view.
\end{definition}%
\nomenclature{$\mathcal{M}_n^{\bullet}([0,1])$}{set of all probability distributions on $\llbracket 1,n \rrbracket^2$}{}{}

We use the cross entropy $\mathbb{H}[\mathbb{P}, \mathbb{Q}] = -\sum_{u,v} \mathbb{P}[u,v]\log_2(\mathbb{Q}[u,v])$ as a measure of similarity on $\mathcal{M}_n^{\bullet}([0,1])$. It can be understood as the expected length of a message generated following $\mathbb{P}$ but encoded with a code optimal for $\mathbb{Q}$. It is minimal when $\mathbb{Q} = \mathbb{P}$, in which case it is equal to the entropy $\mathbb{S}[\mathbb{P}]$. In this paper, the sequence to encode will be $E$, therefore the best compression is achieved for a code based on the empirical distribution $\mathbb{P}_E$.

\begin{figure}[h]
    \centering
    \includegraphics[width=\textwidth]{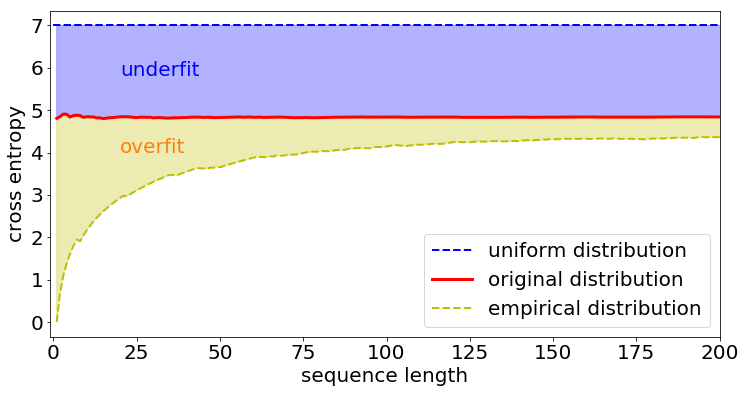}
    \caption{Given an original probability distribution $\mathbb{P}_0$, we generate a sequence of edges $(e_1, \dots, e_m)$. For $k \in \llbracket 1, m \rrbracket$, we plot the cross entropy of the empirical distribution $\mathbb{P}_{(e_1, \dots, e_k)}$ with the uniform distribution (blue line), original distribution (red line), and the empirical distribution itself (yellow line), against $k$. We say that an estimation $\mathbb{Q}^*(e_1, \dots, e_k)$ is overfitting if $\mathbb{H}(\mathbb{P}_{(e_1, \dots, e_k)}, \mathbb{Q}^*(e_1, \dots, e_k))$ lies in the yellow zone, and that it is underfitting if it lies in the blue zone. \label{cross_ent_vs_seq_len}}
\end{figure}

\begin{definition}
Let $\mathbb{P}_E$ be the \emph{empirical distribution}
\[ \forall (u,v) \in \llbracket 1,n \rrbracket^2, \mathbb{P}_E[u,v] = \frac{\#\{k \mid e_k = u \rightarrow v\}}{m}.\]
\end{definition}%
\nomenclature{$\mathbb{P}_0$}{original model distribution}{}{}%
\nomenclature{$\mathbb{P}_E$}{empirical distribution}{}{}%
\nomenclature{$\mathbb{Q}^*(E)$}{estimation of $\mathbb{P}_0$ based on $E$}{}{}%
We can observe that this naive estimation leads to overfitting, as the corresponding code would probably perform poorly for another sequence $E'$ generated using the same original distribution $\mathbb{P}_0$.
On the other hand, the most general code, which performs equally well on all possible edge sequences, is obtained based on the uniform distribution $\mathbb{P}_U$, but it is clearly  underfitting as this code does not tell us anything about $\mathbb{P}_0$. This is illustrated on Figure \ref{cross_ent_vs_seq_len}.
\nomenclature{$\mathbb{S}[\mathbb{P}]$}{entropy of $\mathbb{P}$}{}{}
\nomenclature{$\mathbb{H}[\mathbb{P}, \mathbb{Q}]$}{cross entropy of $\mathbb{P}$ and $\mathbb{Q}$}{}{}

\begin{definition}
    We say that an estimation $\mathbb{Q}^*(E)$ of $\mathbb{P}_0$ is \emph{overfitting} if
    \[ \mathbb{H}[\mathbb{P}_E, \mathbb{Q}^*(E)] < \mathbb{H}[\mathbb{P}_E, \mathbb{P}_0]\]
    on the other hand, we say it is \emph{underfitting} if
    \[ \mathbb{H}[\mathbb{P}_E, \mathbb{Q}^*(E)] > \mathbb{H}[\mathbb{P}_E, \mathbb{P}_0]\]
\end{definition}

For a given sequence $E$, it is very likely that our estimation $\mathbb{Q}^*(E)$ will be at least slightly overfitting or underfitting, but our objective is that 

\begin{equation*}
    \frac{\underset{e_i \sim \mathbb{P}_0}{\mathbb{E}}\left[\mathbb{H}[\mathbb{P}_{(e_1, \dots, e_m)}, \mathbb{Q}^*(e_1, \dots, e_m)]\right]}{\underset{e_i \sim \mathbb{P}_0}{\mathbb{E}}\left[\mathbb{H}[\mathbb{P}_{(e_1, \dots, e_m)}, \mathbb{P}_0]\right]} \underset{m \rightarrow \infty}{\longrightarrow} 1
\end{equation*}

The main risk of overfitting comes from the fact that estimating $\mathbb{Q}^*(E)$ implies the inference of $n^2 - 1$ parameters: we infer $\mathbb{Q}^*(E)[u,v]$ for each $(u,v) \in \llbracket 1,n \rrbracket^2$, under the constraint that $\phi_0(\mathbb{Q}^*(E)) = \sum_{u,v} \mathbb{Q}^*(E)[u,v] - 1 = 0$. As $m$ is typically much smaller than $n^2$, such a large number of parameters induces a high risk of overfitting. 

To avoid this phenomenon, we need to make assumptions about $\mathbb{P}_0$ in order to restrict the search space. We do so by introducing hyperparameters to control the number of degrees of freedom of the model by adding constraints on the probability distribution. A hyperparameter can be described as a function
\begin{align*}
    \phi \colon M_n(\mathbb{R}) &\to \mathbb{R}^{s+1}\\
            \mathbb{Q} &\mapsto (\phi_0(\mathbb{Q}), \dots, \phi_s(\mathbb{Q})
\end{align*}%
\nomenclature{$\phi$}{constraint function restricting the research space}{}{}%
with $\phi_0(\mathbb{Q}) = \sum_{u,v} \mathbb{Q}[u,v] - 1$ is the basic constraint assuring that $\mathbb{Q}$ belongs to $\mathcal{M}_n^{\bullet}([0,1])$ and $s$ is the number of additional constraints. The search space under these constraints is reduced to:
\[ \mathcal{M}_n^{\phi}([0,1]) = \{ \mathbb{Q} \in \mathcal{M}_n([0,1]) \mid \phi(\mathbb{Q}) = 0 \}\]%
\nomenclature{$\mathcal{M}_n^{\phi}([0,1])$}{Subspace of $\mathcal{M}_n^{\bullet}([0,1])$ under the constraint $\phi(\mathbb{Q}) = 0$}{}{}%
We can suppose that the constraints are independent (if they are not, it means that the same search space could be obtained with less constraints). Thus, the number of parameters to infer boils down to $n^2 - s - 1$.

\begin{center}\fbox{\begin{minipage}{0.95\textwidth}
\begin{example}
Let's assume that $\mathbb{P}_0$ is a stochastic blockmodel based on a partition $B=(b_1, \dots, b_p)$. According to the definition given above, it means that 
\[ \exists M \in \mathcal{M}_p([0,1]), \forall u \in b_i, v \in b_j, \mathbb{P}_0[u,v] = M[i,j] \]
It is equivalent to say that
\[ \forall i,j \in \llbracket 1, p \rrbracket, \forall u, u' \in b_i, \forall v, v' \in b_j, \mathbb{P}_0[u,v] - \mathbb{P}_0[u',v'] = 0 \]
which corresponds to a system of $n^2 - p^2$ linearly independent constraints. Thus, under this assumption, we are left with only $p^2 - 1$ parameters to infer. 
\end{example}
\end{minipage}}
\end{center}

Therefore, edge statistical model selection involves two distinct issues:
\begin{enumerate}
    \item For each possible hyperparameter $\phi$, estimate the probability distribution $\mathbb{Q}_{\phi}^*(E)$ that most likely generated $E$ in $\mathcal{M}_n^{\phi}([0,1])$.
    \item Select the best model $\mathbb{Q}^*(E)$ among all possible estimate $(\mathbb{Q}_{\phi}^*(E))_{\phi}$.
\end{enumerate}%
\nomenclature{$\mathbb{Q}_{\phi}^*(E)$}{estimation of $\mathbb{P}_0$ under the constraint $\phi(\mathbb{Q}) = 0$}{}{}
These two main questions are discussed in the next section.

\section{Edge statistical model selection}
\label{model_selection}
\subsection{Parameter inference by minimum description length}
\label{param_selec}

Let's consider first the issue of estimating the probability distribution $\mathbb{Q}_{\phi}^*(E)$ that most likely generated $E$ in $\mathcal{M}_n^{\phi}([0,1])$, given the hyperparameter $\phi$. We rely on the minimum description length principle (a detailed tutorial can be found in \cite{grunwald2005minimum}). It states that, as any regularity in a sequence of observations can be used to compress it, the best statistical model for the sequence $E$ is the one which minimizes the description length of the model $D_{\phi}(\mathbb{Q})$ plus the description length of the observations compressed using this model $D(E|\mathbb{Q})$:
\begin{equation}
    \label{mdl}
    \mathbb{Q}_{\phi}^*(E) = \underset{\mathbb{Q} \in \mathcal{M}_n^{\phi}([0,1])}{\mathrm{argmin}} \; D(E|\mathbb{Q}) + D_{\phi}(\mathbb{Q})
\end{equation}%
\nomenclature{$D_{\phi}({\mathbb{Q}})$}{description length of a model $\mathbb{Q}$ within $\mathcal{M}_n^{\phi}([0,1])$}{}{}%
The description length of the sequence can be computed as \[ D(E|\mathbb{Q}) = -\sum_{i=1}^m \log_2(\mathbb{Q}[e_i])\]
as detailed in Annex~\ref{annex:dl}.%
\nomenclature{$D(E \mid \mathbb{Q})$}{description length of sequence $E$ using model $\mathbb{Q}$}{}{}%

Then, to compute the description length of the model $D_{\phi}(\mathbb{Q})$, we need to define a probability distribution $\bar{\mathbb{P}_{\phi}}$ on $\mathcal{M}_n^{\phi}([0,1])$. This so-called prior distribution is used to encode the model with a length $D_{\phi}(\mathbb{Q}) = -\log_2(\bar{\mathbb{P}_{\phi}}[\mathbb{Q}])$. The goal of this term is to take into account the complexity of the model, in order to avoid overfitting. Therefore, simpler models should have shorter description length. To achieve this, we define the prior distribution such that the description length of a model is inversely proportional to its information content, measured by its entropy:
\[ \bar{\mathbb{P}}_{\phi}[\mathbb{Q}] = \frac{1}{Z_{\phi}} \times 2^{\mathbb{S}[\mathbb{Q}]} \]
with $Z_{\phi} = \int_{\mathcal{M}_n^{\phi}([0,1])} 2^{\mathbb{S}[\mathbb{Q}]} \,\mathrm{d}\mathbb{Q}$ a normalization constant to ensure that $\bar{\mathbb{P}}_{\phi}[\mathbb{Q}]$ integrates to $1$ over $\mathcal{M}_n^{\phi}([0,1])$. 

\begin{center}
\fbox{\begin{minipage}{0.95\textwidth}
\begin{remark}
The expression prior distribution we used to refer to $\bar{\mathbb{P}}_{\phi}$ refers to the bayesian terminology. This is on purpose, as this approach is equivalent to bayesian statistical inference, as detailed in Annex~\ref{annex:bay_inf}.
\end{remark}
\end{minipage}}
\end{center}

With this definition of $\bar{\mathbb{P}}_{\phi}$, 
\begin{align*}
D_{\phi}[\mathbb{Q}] &= -\log_2(\bar{\mathbb{P}}_{\phi}[\mathbb{Q}]) \\
&= - \mathbb{S}[\mathbb{Q}] + \log_2(Z_{\phi})
\end{align*}
As $Z_{\phi}$ is constant on $\mathcal{M}_n^{\phi}([0,1])$, we can neglect it in the minimization and equation \ref{mdl} becomes 
\begin{equation}
\label{mdl2}
    \mathbb{Q}_{\phi}^*(E) = \underset{\mathbb{Q} \in \mathcal{M}_n^{\phi}([0,1])}{\mathrm{argmin}} \; -\sum_{i=1}^m \log_2(\mathbb{Q}[e_i]) - \mathbb{S}[\mathbb{Q}]
\end{equation}
In the following, we denote
\[\mathrm{f}(\mathbb{Q}, E) = - \sum_{i=1}^m \log_2(\mathbb{Q}[e_i]) + \sum_{u,v}\mathbb{Q}[u,v]\log_2(\mathbb{Q}[u,v]) \]
and thus we can rewrite equation \ref{mdl2} as
\begin{equation}
    \mathbb{Q}_{\phi}^*(E) = \underset{\mathbb{Q} \in \mathcal{M}_n^{\phi}([0,1])}{\mathrm{argmin}} \; \mathrm{f}(\mathbb{Q},E)
\end{equation}
We have the following property (see proof in Annex \ref{minimum_proof}):
\begin{center}
\fbox{\begin{minipage}{0.95\textwidth}
\begin{property}
\label{minimum_exist_unique}
If $\mathcal{M}_n^{\phi}([0,1])$ is a convex set, then for any edge sequence $E$, $\mathrm{f}$ has a unique minimum $\mathbb{Q}_{\phi}^*(E)$ over $\mathcal{M}_n^{\phi}([0,1])$.
\end{property}
\end{minipage}}
\end{center}

\begin{remark}
\label{rq1}
In particular, if $\phi$ is an affine function, $\mathcal{M}_n^{\phi}([0,1])$ is the intersection of an affine subspace of $\mathcal{M}_n(\mathbb{R})$ with $[0,1]^{n^2}$. Consequently, it is convex and $\mathbb{Q}_{\phi}^*(E)$ exists and is unique.
\end{remark}

According to the Lagrange multiplier theorem, this minimum verifies
\[ \exists (\lambda_j) \in \mathbb{R}^{s+1}, \vec{\nabla}f(\mathbb{Q}_{\phi}^*(E), E) + \sum_{j = 1}^{s+1} \lambda_j \vec{\nabla}\phi_j(Q_{\phi}^*(E)) = 0\]
This is a set of $n^2 + s + 1$ equations with as many unknowns which we solve numerically using Newton's method.

Finally, we obtain the following result (see proof in Annex~\ref{annex:proofcv}):
\begin{center}
\fbox{\begin{minipage}{0.95\textwidth}
\begin{theorem}
Let $(e_i)_{i \in \mathbb{N}}$ be a sequence of independent and identically distributed random variables following $\mathbb{P}_0 \in \mathcal{M}_n^{\bullet}([0,1])$. 

\[\forall \phi, \mathbb{Q}_{\phi}^*(e_1, \dots, e_x) \underset{x \rightarrow \infty}{\longrightarrow} \underset{\mathbb{Q} \in \mathcal{M}_n^{\phi}([0,1])}{\mathrm{argmin}} \mathbb{H}(\mathbb{P}_0, \mathbb{Q})\]
\end{theorem}
\end{minipage}}
\end{center}

\begin{remark}
In particular, if $\mathbb{P}_0$ belongs to $\mathcal{M}_n^{\phi}([0,1])$, it means that $\mathbb{Q}_{\phi}^*(E)$ converges toward $\mathbb{P}_0$ as the number of observations grows.
\end{remark}

\subsection{Hyperparameter selection by sequential update}


Now that we know how to infer $\mathbb{Q}_{\phi}^*(E)$ for any given $\phi$, the second step for model selection consists to choose the best estimation $\mathbb{Q}^*(E)$ among them. Let's consider a set of hyperparameters $\Phi = \{\phi_1, \dots, \phi_q\}$. To select the best hyperparameter $\phi^*(E) \in \Phi$, we keep using the minimum description length. The difference with parameter selection is that, to compute the description length of $E$ given a hyperparameter $\phi$, we need a training set of edges $L$ on which we can learn $\mathbb{Q}_{\phi}^*(L)$, and use this model to encode $E$ with a length
\begin{align*}
    D[E|\phi] &= -\sum_{i = 1}^{m}\log_2(\mathbb{Q}_{\phi}^*(L)[e_i]) \\
        &= m \times \mathbb{H}[\mathbb{P}_E, \mathbb{Q}_{\phi}^*(L)]
\end{align*}
$L$ is necessarily a subset of $E$, but the issue is its size. The smaller it is, the more we risk underfitting: the extreme example is for $L=\emptyset$, because then $\forall \phi, \mathbb{Q}_{\phi}^*(L)$ is the uniform distribution, which is the extreme case of underfitting. On the other hand, the larger the size of the learning set, the more we favour hyperparameters with many degrees of freedom and risk overfitting: typically, if $L=E$, $D[E|\phi]$ is smaller as $\mathbb{P}_E \in \mathcal{M}_n^{\phi}([0,1])$ and $\mathbb{Q}_{\phi}^*(E)$ is close to $\mathbb{P}_E$. 

To avoid both pitfalls, instead of a fixed learning set, we use sequential learning. Let's consider the situation where $E$ is a sequence of messages that a source (Alice) draws at random and transmits to a destination (Bob). Instead of using a fixed code $C^*(L)$, Alice updates her code as she observes more and more edges. At step $k$, Alice has observed edges $(e_1, \dots, e_{k-1})$ and she has transmitted them to Bob. Therefore, both of them can compute $\mathbb{Q}_{\phi}^*(e_1, \dots, e_{k-1})$ and the corresponding code $C_{\phi}^*(k-1)$. Alice draws the edge $e_k$ and transmits it to Bob using this code, then Alice and Bob both update their code to $C_{\phi}^*(k)$, and so on. This way, the description length is

\[ D[E|\phi] = - \sum_{k = 1}^m \log_2(\mathbb{Q}_{\phi}^*(e_1, \dots, e_{k-1})[e_k]) \]

At step $k$, the probability distribution $\mathbb{Q}_{\phi}^*(e_1, \dots, e_{k-1})$ is the model that best fits the first $k-1$ observations within $\mathcal{M}_n^{\phi}([0,1])$. Thus, $\mathbb{Q}_{\phi}^*(e_1, .., e_{k-1})[e_k]$ is the probability, given those observations and the hyperparameter, to correctly guess the $k^{th}$ edge, and $-\log_2(\mathbb{Q}_{\phi}^*(e_1, \dots, e_{k-1})[e_k])$ can be interpreted as the quantity of information about $e_k$ contained in the previous edges. The constraints imposed by the hyperparameter $\phi$ induce correlations and allow to predict the appearance of edges not yet observed. Therefore, choosing the right hyperparameter is equivalent to choose the right level of constraint. It has to be low enough to adapt to observations, but high enough to predict unobserved edges. We consider a uniform distribution on $\Phi$ as prior distribution, thus $D(\Phi)$ is constant and the best hyperparameter is computed as

\[ \phi^*(E) = \underset{\phi \in \Phi}{\mathrm{argmin}} \; - \sum_{k = 1}^m \log_2(\mathbb{Q}_{\phi}^*(e_1, \dots, e_{k-1})[e_k]) \]

Overall, the model selected is

\[ \mathbb{Q}^*(E) = \mathbb{Q}^*_{\phi^*(E)}(E) \]

Sequential update implies that the optimal model $\mathbb{Q}^*(E)$ is dependent on the order of edges in $E$. This means that if the studied graph $G$ is static, the model selected depends on the ordering of edges we make when we choose a representative $E \in \mathcal{E}^{\downarrow}_G$. However, we observe in practice that various edge ordering have little impact on the results.


\section{Applications to model selection}
\label{application to model selection}
\subsection{Stochastic blockmodel partition selection}
\label{sbm partition}
\paragraph{Finding the appropriate number of blocks of the partition.}
To test sequential edge probability inference, we start by using it to tackle the classical problem of partition selection in stochastic blockmodels. We consider an edge sequence $E = (e_1, \dots, e_m)$ which we assume was generated by a stochastic blockmodel $\mathbb{P}_0$ based on a partition $B_0$, as described in section \ref{sbm_def}. Our objective is to retrieve $\mathbb{P}_0$ and $B_0$ among the set of all possible stochastic blockmodels. Each partition $B = (b_1, \dots, b_p)$ of $\llbracket 1, n \rrbracket$ corresponds to a hyper\-para\-meter $\phi^B$ made of $n^2 - p^2 + 1$ constraints. If we designate inside each block $b_i$ a representative $u_i$, this hyperparameter can be expressed as:
\begin{gather*}
\phi^B_0(\mathbb{Q}) = 1 - \sum_{u,v} \mathbb{Q}[u,v] \\
\forall i,j, \forall u \in b_i \setminus \{u_i\}, \forall v \in b_j \setminus \{u_j\}, \phi^B_{u,v}(\mathbb{Q}) = \mathbb{Q}[u_i, u_j] - \mathbb{Q}[u,v] 
\end{gather*}

The constraint $\phi^B(\mathbb{Q}) = 0$ expresses the fact that $\mathbb{Q}$ is a probability distribution and that edge generation probabilities are constant along the blocks defined by $B$. Thus, selecting the partition within a set $\{B_1, \dots, B_q\}$ that is more likely to be the original one boils down to the inference of the most likely hyperparameter in $\Phi = \{\phi^{B_1}, \dots, \phi^{B_q}\}$. In particular, it should be noted that all those hyperparameters are affine functions, so Remark~\ref{rq1} 
tells us that for each of them, $\mathbb{Q}_{\phi}^*(E)$ exists, is unique, and can be computed using Lagrange multipliers and Newton's method.

Exploring the full partition space is a challenge on its own, as this space grows exponentially with $n$. Therefore, to perform our test, we generate synthetic graphs with a stochastic blockmodel and observe how it behaves for a particular subset of the possible partitions of the nodes. Of course, this means that we cannot be sure that the minimum we find corresponds to the minimum over every possible partition. Yet, it allows us to test the robustness of sequential edge probability inference against common pitfalls, and in particular with respect to partitions which are a coarsening or a refinement of the original partition.

We consider a stochastic blockmodel $S_0 = (B_0,M_0)$ on $128$ nodes divided into $4$ blocks:
\begin{gather*}
    B = \llbracket 1, 32 \rrbracket, \llbracket 33, 64 \rrbracket, \llbracket 65, 96 \rrbracket, \llbracket 97, 128 \rrbracket \\
    M = \frac{1}{128^2} \cdot \begin{bmatrix} 4 & 0 & 0 & 0 \\ 0 & 4 & 0 & 0 \\ 0 & 0 & 4 & 0 \\ 0 & 0 & 0 & 4 \end{bmatrix}
\end{gather*}

\begin{remark}
    As the stochastic blockmodel defined here is an edge statistical model, the coefficients $M[i,j]$ should not be interpreted as the density between blocks $i$ and $j$. They are the probability for each edge going from block $i$ to block $j$ to be generated:
    \[ \forall u \in b_i, v \in b_j, \mathbb{P}_{S_0}[u,v] = M[i,j] \]
\end{remark}

\begin{figure}[h]
    \centering
    \scalebox{1}{\includegraphics[width=\textwidth]{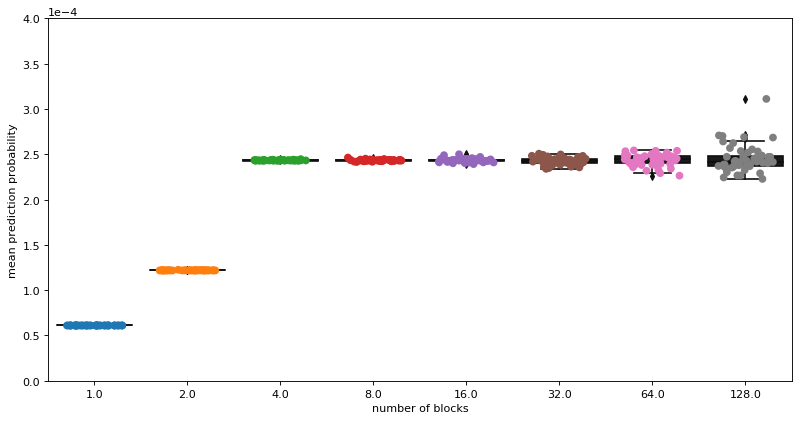}}
    \caption{For each of the $50$ graphs generated with $S_0 = (B_0,M_0)$, we plot the mean prediction probability against the number of blocks of the hyperparameter for partitions ranging from $1$ single block to $128$ blocks containing a single node. \label{mean_prob_vs_nb_com}}
    \end{figure}

We generate $50$ graphs with $S_0$ and test $8$ hyperparameters corresponding to partitions refined from $1$ block to $128$. Each partition is obtained by dividing the blocks of the previous one in half.
We plot the mean prediction probability $\frac{1}{m} \sum_{k=1}^{m} \mathbb{Q}^*_{\phi^ B}(e_1, \dots, e_{k-1})[e_k]$ against the number of blocks in $B$. Results are shown in Figure \ref{mean_prob_vs_nb_com}. We observe that the mean prediction probability rises as the number of blocks of the partition grows from one to four, which corresponds to the original partition used to generate the graphs. Then, further refinement of the partition used as hyperparameter does not bring significant increase in the mean prediction probability.

    \begin{figure}[h]
    \centering
    \includegraphics[width=0.98\textwidth]{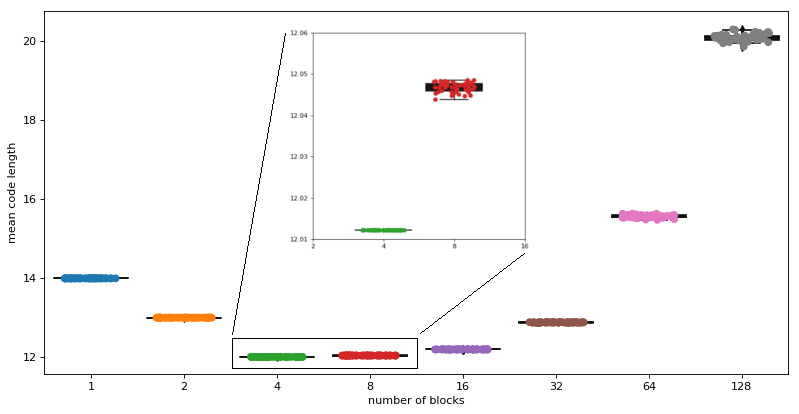}
    \caption{For the $50$ graphs generated with $S_0$, and the eight partitions obtained by coarsening / refining $B_0$, we plot the mean code length against the number of blocks in the partition. \label{mean_code_len_vs_nb_com}}
\end{figure}

Then, we plot the mean code length $-\frac{1}{m} \sum_{k=1}^{m} \log_2(\mathbb{Q}^*_{\phi^B}(e_1, \dots, e_{k-1})[e_k])$ against number of blocks in $B$ in Figure \ref{mean_code_len_vs_nb_com}. The mean code length is proportional to the description length $D[E|\phi^B]$ so they have the same minimum, but it has the advantage of being insensitive to the length of the edge sequence. We observe that for all fifty graphs, it presents a clear minimum at the original four blocks partition $B_0$. For coarser partitions, the mean code length is higher because, as illustrated in Figure \ref{mean_prob_vs_nb_com}, the prediction probability is lower and
\begin{align*}
    \mathbb{Q}^*_{\phi^B}(e_1, \dots, e_{k-1})[e_k] &< \mathbb{Q}^*_{\phi^{B_2}}(e_1, \dots, e_{k-1})[e_k] \\ \implies -\log_2(\mathbb{Q}^*_{\phi^B}(e_1, \dots, e_{k-1})[e_k]) &> -\log_2(\mathbb{Q}^*_{\phi^{B_2}}(e_1, \dots, e_{k-1})[e_k])
\end{align*}
Then, for finer partitions, it is due to the slower convergence rate. Indeed, as logarithm is a concave function
\[-\log_2\left(\frac{1}{m} \sum_{k=1}^{m} \mathbb{Q}^*_{\phi^B}(e_1, \dots, e_{k-1})[e_k]\right) < -\frac{1}{m} \sum_{k=1}^{m} \log_2(\mathbb{Q}^*_{\phi^B}(e_1, \dots, e_{k-1})[e_k]) \]
Therefore, the greater the fluctuations of the prediction probability, the higher the mean code length. More details about the convergence of the prediction probability depending on the hyperparameter can be found in Annex \ref{edge_prediction_probability}. At the end, the minimum description length makes it possible to retrieve the original partition $B_0$, avoiding both overfitting and underfitting, with no previous knowledge or assumption about the number of blocks.

\paragraph{Merge / split issue.}
It has been shown in \cite{duvivier2019minimum} that stochastic blockmodel selection based on the minimization of the microcanonical ensemble entropy, even though they are statistically grounded, may be subject to overfitting in the sense that splitting large communities while merging small ones may lead to a lower entropy because it imposes more constraints on edges' position.

To illustrate how sequential edge probability inference helps solving this problem, let's consider a stochastic blockmodel $S_1 = (B,M)$ defined on a set of $n = 12$ nodes: 
\begin{gather*}
   B = \llbracket0;5\rrbracket, \llbracket6;8\rrbracket, \llbracket9;11\rrbracket \\
   M = \begin{bmatrix}
  0.026 & 0 & 0 \\
  0 & 0.003 & 0 \\
  0 & 0 & 0.003
  \end{bmatrix} 
\end{gather*}

We test two different partitions: the original one, $B$, and the inverse partition in which the large communities is split and small ones are merged $B^{\dagger} = \llbracket0;2\rrbracket, \llbracket3;5\rrbracket, \llbracket6;11\rrbracket$. To do so, we generate $100$ graphs $G_i$ made of $m = 378$ edges with $S_1$ and for each graph, we compute the mean code length and the entropy (using graphtools\footnote{\url{https://graph-tool.skewed.de}}) for both partitions. Then, for both quality functions, we compute the percentage of graphs for which the original partition is identified as better than the inverse one. Results are shown in Table \ref{correct_match}. While the mean code length almost always correctly identifies the original partition, the entropy of the microcanonical ensemble never does so. The graphs considered here have a very high density, which makes them not very realistic, but same results can be obtained with lower density graphs. Let's consider a stochastic blockmodel $S_2$ with $n = 256$ nodes, partitioned in $33$ communities, one of size $128$, and $32$ of size $4$: 
\[B = \llbracket 1, 128 \rrbracket, \llbracket 129, 132 \rrbracket, \llbracket 133, 136 \rrbracket, \dots, \llbracket 253, 256 \rrbracket\]
The internal probability of the big community is $6\times 10^{-5}$, the one of the small communities is $7.6\times 10^{-4}$, and the probability between communities is null. We compare this original partition with the inverse one:
\[B^{\dagger} = \llbracket 1, 4 \rrbracket, \llbracket 5, 8 \rrbracket, \dots, \llbracket 125, 128 \rrbracket, \llbracket 129, 256 \rrbracket\]
We generate $100$ graphs with $S_2$ and compute for each of them the entropy of both partitions and the mean code length with $\phi_B$ and $\phi_{B^{\dagger}}$. Results are shown in Table~\ref{correct_match}. We see that in this case too, the mean code length always identifies the original partition as the best one, while the entropy does not.

\begin{table}
  \begin{center}
    \caption{Percentage of correct match for heterogeneous graphs.\label{correct_match}}
    \begin{tabular}{c|c|c}
    SBM & Mean code length & Entropy \\
    \hline
    $S_1$ & $96\%$ & $0\%$ \\
    $S_2$ & $100\%$ & $0\%$ 
    \end{tabular}
  \end{center}
\end{table}

A more thorough investigation of how sequential edge probability inference applies to the specific case of stochastic blockmodels can be found in \cite{duvivier2020edge}.

\subsection{Stochastic blockmodel and configuration model}
\label{sbm and cfm}
The main benefit of edge statistical models is that it provides a common framework to compare models of different natures. To illustrate this, let's consider two widespread models: the stochastic blockmodel and the configuration model. The first one has been introduced in the previous section, so we  start by describing the edge-version of the configuration model, and then show how both models can be compared using the minimum description length.

We consider the directed version of the configuration model. The classical version of this model takes as parameters the sequences of node in  $(k^{out}_u)_{u \in V}$ and out $(k^{in}_u)_{u \in V}$ degrees. For the edge version, we keep the idea that the probability of generating an edge $u \rightarrow v$ is determined by two probability distributions $p^{out}$ and $p^{in}$ over $\llbracket 1,n \rrbracket$. $p^{out}_u$ is the probability to pick node $u$ as the source of the edge and $p^{in}_v$ the probability to pick $v$ as its destination:
\[ \forall u,v, \mathbb{Q}_{CM}[u,v] = p^{out}_u \times p^{in}_v \]
Therefore, a probability distribution $\mathbb{Q} \in \mathcal{M}_n^{\bullet}([0,1])$ corresponds to a directed configuration model if and only if:
\[ \forall u,v, \mathbb{Q}[u,v] \times \mathbb{Q}[1,1] - \mathbb{Q}[u,1] \times \mathbb{Q}[1,v] = 0\]
In this case, $p^{out}_u = \sum_v \mathbb{Q}[u,v]$ and $p^{in}_v = \sum_u \mathbb{Q}[u,v]$. This gives us a system of $(n-1)^2$ independent constraints to use as hyperparameter $\phi^{CM}$. It is worth noting that this hyperparameter is not an affine function, so Remark~\ref{rq1} does not apply. However, we have the following result (see proof in Annex~\ref{minimum_cfm}):
\begin{center}
\fbox{\begin{minipage}{0.95\textwidth}
\begin{property}
\label{minimum_exist_unique_cfm}
For any edge sequence $E$, $\mathrm{f}$ has a unique minimum $\mathbb{Q}_{\phi^{CM}}^*(E)$ over $\mathcal{M}_n^{\phi^{CM}}([0,1])$.
\end{property}
\end{minipage}}
\end{center}

Yet, this still leaves $2n - 2$ parameters to infer, which remains high in comparison with the number of observations $m$ and thus induces a risk of overfitting. To overcome this problem, we consider a block version of the configuration model. It means that, given two partitions of $\llbracket 1,n \rrbracket$, $B^{in}$ and $B^{out}$, $(p^{in}_u)_{u \in \llbracket 1,n \rrbracket}$ is constant over the blocks of $B^{in}$ and $(p^{out}_u)_{u \in \llbracket 1,n \rrbracket}$ is constant over the blocks of $B^{out}$. Thus, if $B^{in}$ is made of $q^{in}$ blocks and $B^{out}$ of $q^{out}$ blocks, there are only $q^{out} + q^{in} - 2$ parameters left to infer.

In practice, we consider two models on $n=128$ nodes: the stochastic blockmodel $S_1 = (B_1, M_1)$ (and its associated hyperparameter $\phi_{B_1}$) defined as \begin{gather*}
    B_1 = \llbracket 1, 64 \rrbracket, \llbracket 65, 128 \rrbracket \\
    M_1 = \frac{1}{n^2} \cdot \begin{bmatrix} 2 & 0 \\ 0 & 2 \end{bmatrix}
\end{gather*}
and the block configuration model $CM$ defined by
\begin{gather*}
    B^{out} = \llbracket 1, 96 \rrbracket, \llbracket 97, 120 \rrbracket, \llbracket 121, 126 \rrbracket, \llbracket 127, 128 \rrbracket \\
    p^{out} = [0.0054; 0.0109; 0.0217; 0.0435] \\
    B^{in} = (\llbracket 1, 2 \rrbracket, \llbracket 3, 8 \rrbracket, \llbracket 9, 32 \rrbracket, \llbracket 33, 128 \rrbracket) \\
    p^{in} = [0.0435; 0.0217; 0.0109; 0.0054]
\end{gather*}
which corresponds to a hyperparameter $\phi_{B^{out}, B^{in}}$.

\begin{figure}[ht!]
    \centering
    \includegraphics[width=\textwidth]{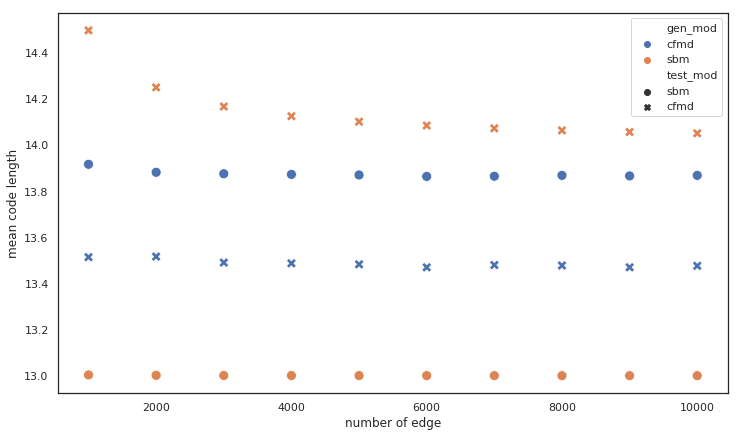}
    \caption{Mean code length of two families of edges sequences, encoded using stochastic blockmodel hyperparameter and configuration model hyperparameter.}
    \label{mcl_vs_m}
    \end{figure}

For each probability distribution $\mathbb{P}_{S_1}$ and $\mathbb{P}_{CM}$, we generate $10$ edge sequences of length $1000$ to $10000$. Then, for each edge sequence, we compute its mean code length using hyperparameters $\phi_{B_1}$ and $\phi_{B^{out}, B^{in}}$. Results are shown in Figure \ref{mcl_vs_m}.

    \begin{figure}[ht!]
    \includegraphics[width=\textwidth]{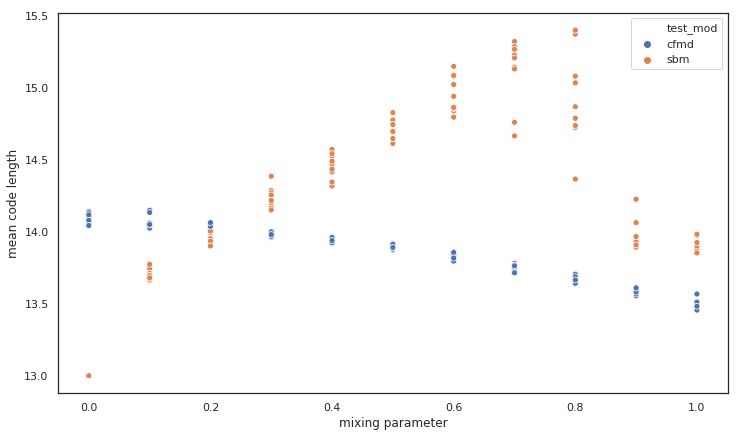}
    \caption{Mean code length against mixing parameter.}
    \label{mcl_vs_l}
\end{figure}

We observe that, for the sequences of edges which are generated using $\mathbb{P}_{CM}$ (blue dots and crosses), the mean code length is lower when using the configuration model hyperparameter $\phi_{B^{out}, B^{in}}$. On the other hand, for the sequences generated using $\mathbb{P}_{S_1}$ (yellow dots and crosses), the mean code length is lower when using the stochastic blockmodel hyperparameter $\phi_{B_1}$. Thus, the best compression actually corresponds to the correct hyperparameter.

What is even more interested, is that we can also use sequential edge probability inference to identify the most significant property when the edge distribution is the result of a combination of factors. Continuing with models $\mathbb{P}_{S_1}$ and $\mathbb{P}_C$, let's define the mixed model:
\[ \mathbb{P}(\lambda) = \lambda \cdot \mathbb{P}_C + (1 - \lambda) \cdot \mathbb{P}_{S_1}\]
We consider $11$ values of $\lambda$ between $0$ and $1$, and for each, we generate $10$ edge sequences of length $2800$. Then, for each edge sequence, we compute its mean code length using $\phi_{B^{out}, B^{in}}$ and $\phi_{B_1}$. Results are shown in Figure \ref{mcl_vs_l}


We observe that as $\lambda$ rises from $0$ to $1$, the mean code length using the block hyperparameter $\phi_{B_1}$ rises from $13$ to $14$, with a pick up to $15.5$. On the other hand, the mean code length using the configuration structure decreases from a little more than $14$ down to $13.5$. It shows that the mean code length is able to capture the increasing influence of the block structure and the decreasing influence of the configuration structure in the distribution of edges. When one model clearly dominates the other (\textit{i.e.} $\lambda \leq 0.2$ or $\lambda \geq 0.8$) the corresponding hyperparameter leads to a better compression. 


\section{Conclusion}
\label{sec:conclusion} 
In conclusion, we have introduced sequential edge probability inference, a new statistical framework to perform model selection on graphs. Describing models of various nature as probability distributions on edge allows to easily compare their performance thanks to minimum description length (or equivalently bayesian inference). Moreover, by introducing additional constraints as hyperparameters, we are able to lower the number of parameters of the model below the number of observations on which inference is performed, which is necessary to avoid overfitting.

We have illustrated how this framework can be used to select the most significant node partition according to information present in edge distribution. Because it relies on statistical inference, it provides a simple way to discriminate automatically between too fine and too coarse partitions with no a priori information.

The main advantage of sequential edge probability inference is that it provides a common formulation of models of different nature in order to compare them. It is thus able, for example, to automatically detect whether the distribution of edges is determined rather by nodes' block membership (block structure) or by their potential to emit or receive edges (configurational structure), even in cases where both structures are mixed.

We believe these results to be only a foretaste of the potential of this approach. Because it has firm theoretical grounds, we are convinced that it can provide fruitful applications in many domains where interactions are the results of entangled mechanisms whose effect on the overall graph topology can only be told apart by rigorous statistical analysis. It therefore provides a reliable criterion which, combined with a methodology to explore the hyperparameter search space, can lead to the automatic selection of the best model for a given graph.

\bibliographystyle{unsrt}
\bibliography{biblio}

\clearpage
\section{Annex}
\subsection{Proof of existence and unicity of the minimum}
\label{minimum_proof}
We prove the following result
\begin{center}
\fbox{\begin{minipage}{0.95\textwidth}
If $\mathcal{M}_n^{\phi}([0,1])$ is a convex set, then for any edge sequence $E$, $\mathrm{f}$ has a unique minimum $\mathbb{Q}_{\phi}^*(E)$ over $\mathcal{M}_n^{\phi}([0,1])$.
\end{minipage}}
\end{center}
Let's consider $\phi$ such that $\mathcal{M}^{\phi}_n([0,1])$ is a convex set, and let $E$ be an edge sequence. Let's denote
\[ \mathcal{M}^{\phi}_n(]0,1]) = \{ \mathbb{Q} \in \mathcal{M}^{\phi}_n([0,1]) \mid \forall u,v, \mathbb{Q}[u,v] > 0\}\]
All $\mathbb{Q}$ thus removed from $\mathcal{M}^{\phi}_n([0,1])$ lies on its boundary, so as it is supposed to be convex, $\mathcal{M}^{\phi}_n(]0,1])$ is convex too. We consider the function
\begin{align*}
    \mathrm{f}_E : \mathcal{M}^{\phi}_n(]0,1]) &\to \mathbb{R} \\
    \mathbb{Q} &\mapsto - \sum_{i=0}^{m-1} \log_2(\mathbb{Q}[e_i]) + \sum_{u,v}\mathbb{Q}[u,v]\log_2(\mathbb{Q}[u,v])
\end{align*}
For all pairs of nodes $(u,v)$, we denote 
\[ K_{u,v} = \#\{k \in \llbracket 0,m-1 \rrbracket \mid e_k = u \rightarrow v\}\]
Then, $\mathrm{f}_E$ can be rewriten
\begin{align*} 
\forall \mathbb{Q}, \mathrm{f}_E(\mathbb{Q}) &= \sum_{u,v \in \llbracket 0, n-1 \rrbracket} (\mathbb{Q}[u,v] - K_{u,v})\log_2(\mathbb{Q}[u,v])
\end{align*}
$\mathrm{f}_E$ is $\mathcal{C}_2$ on $\mathcal{M}^{\phi}_n(]0,1])$ and it's Hessian matrix is
\[
\begin{bmatrix} \frac{K_{0,0}}{\mathbb{Q}[0,0]^2} + \frac{1}{\mathbb{Q}[0,0]} & 0 & \dots & 0 \\ 0 & \frac{K_{1,0}}{\mathbb{Q}[1,0]^2} + \frac{1}{\mathbb{Q}[1,0]} & \dots & 0 \\ \vdots & \vdots & \ddots & 0 \\ 0 & 0 & 0 & \frac{K_{n-1,n-1}}{\mathbb{Q}[n-1,n-1]^2} + \frac{1}{\mathbb{Q}[n-1,n-1]} \end{bmatrix}
\]
which is positive definite on $\mathcal{M}^{\phi}_n(]0,1])$, so $\mathrm{f}_E$ is strictly convex on this set. As $\mathcal{M}^{\phi}_n(]0,1])$ is a convex set, we obtain that $\mathrm{f}_E$ has a unique minimum over it, which we can denote $\mathbb{Q}_{\phi}^*(E)$. It remains to be proven that $\mathbb{Q}_{\phi}^*(E)$ is the minimum of $\mathrm{f}_E$ over $\mathcal{M}^{\phi}_n([0,1])$. $\mathcal{M}^{\phi}_n([0,1])$ is the closure (in the topological sense) of $\mathcal{M}^{\phi}_n(]0,1])$, so $\mathrm{f}_E$ can be continuously extended to it, provided that we extend it's codomain to $\bar{\mathbb{R}} = \mathbb{R} \cup \infty$. Let's consider $\mathbb{Q} \in \mathcal{M}_n^{\phi}([0,1])$ such that $\exists u,v, \mathbb{Q}[u,v] = 0$ and a sequence $(\mathbb{Q}_i)_{i \in \mathbb{N}} \in \mathcal{M}_n^{\phi}(]0,1])$ that converges toward $\mathbb{Q}$. There are two different situations.

\paragraph{1.} If $\exists u_0,v_0, \mathbb{Q}[u_0,v_0] = 0 \land K_{u_0,v_0} > 0$. Then, 
\begin{align*}
    \forall i, \mathrm{f}_E(\mathbb{Q}_i) 
    = &\sum_{u,v \in \llbracket 0, n-1 \rrbracket} (\mathbb{Q}_i[u,v] - K_{u,v})\log_2(\mathbb{Q}_i[u,v]) \\
    = &\sum_{u,v \neq u_0, v_0} (\mathbb{Q}_i[u,v] - K_{u,v})\log_2(\mathbb{Q}_i[u,v]) + \\ &(\mathbb{Q}_i[u_0,v_0] - K_{u_0, v_0})\log_2(\mathbb{Q}_i[u_0,v_0])
\end{align*}
Thus,
\[
    \mathrm{f}_E(\mathbb{Q}_i) \underset{i \rightarrow \infty}{\longrightarrow} \infty
\]
So we define $\mathrm{f}_E(\mathbb{Q}) = \infty$ and in particular $\mathrm{f}_E(\mathbb{Q}) > \mathrm{f}_E(\mathbb{Q}^*_{\phi}(E))$.

\paragraph{2.} If $\forall u,v, \mathbb{Q}[u,v] = 0 \Rightarrow K_{u,v} = 0$. Then, 
\begin{align*}
    \forall i, \mathrm{f}_E(\mathbb{Q}_i) 
    = &\sum_{u,v \in \llbracket 0, n-1 \rrbracket} (\mathbb{Q}_i[u,v] - K_{u,v})\log_2(\mathbb{Q}_i[u,v]) \\
    = &\sum_{u,v \mid \mathbb{Q}[u,v] > 0} (\mathbb{Q}_i[u,v] - K_{u,v})\log_2(\mathbb{Q}_i[u,v]) + \\  &\sum_{u,v \mid \mathbb{Q}[u,v] = 0} \mathbb{Q}_i[u,v]\log_2(\mathbb{Q}_i[u,v])
\end{align*}
Thus,
\[
    \mathrm{f}_E(\mathbb{Q}_i) \underset{i \rightarrow \infty}{\longrightarrow} \sum_{u,v \mid \mathbb{Q}[u,v] > 0} (\mathbb{Q}[u,v] - K_{u,v})\log_2(\mathbb{Q}[u,v])
\]
and we define
\[
    \mathrm{f}_E(\mathbb{Q}) = \sum_{u,v \mid \mathbb{Q}[u,v] > 0} (\mathbb{Q}[u,v] - K_{u,v})\log_2(\mathbb{Q}[u,v])
\]
By continuity of $\mathrm{f}_E$, we know that $\mathrm{f}_E(\mathbb{Q}) \geq \mathrm{f}_E(\mathbb{Q}_{\phi}^*(E))$. Let's show that this inequality is strict. We consider the restriction of $\mathrm{f}_E$ to the interval 
\[I = \{\lambda \cdot \mathbb{Q}_{\phi}^*(E) + (1 - \lambda)\cdot\mathbb{Q}, \lambda \in [0, 1[\} \subset \mathcal{M}_n^{\phi}(]0,1])\]
Because of the strict convexity of $\mathrm{f}_E$ on $\mathcal{M}_n^{\phi}(]0,1])$, $\mathrm{f}_E\big|_I$ is a strictly increasing function of $\lambda$. As a consequence, 
\[
\mathrm{f}_E(\mathbb{Q}_{\phi}^*(E)) < \underset{\lambda \rightarrow 1}{\mathrm{lim}} \: \mathrm{f}_E\big|_I(\lambda \cdot \mathbb{Q}_{\phi}^*(E) + (1 - \lambda)\cdot\mathbb{Q}) = \mathrm{f}_E(\mathbb{Q})
\]
Which proves that in both cases, $\mathbb{Q}_{\phi}^*(E)$ is the only minimum of $\mathrm{f}$ over $\mathcal{M}_n^{\phi}([0,1])$.
\subsection{Proof of existence and unicity of the minimum (configuration model)}
\label{minimum_cfm}
We prove the following result
\begin{center}
\fbox{\begin{minipage}{0.95\textwidth}
For any edge sequence $E$, $\mathrm{f}$ has a unique minimum $\mathbb{Q}_{\phi^{CM}}^*(E)$ over $\mathcal{M}_n^{\phi^{CM}}([0,1])$.
\end{minipage}}
\end{center}
Let's define the set of probability distributions on $\llbracket 1, n \rrbracket$:
\begin{definition}
    We denote $\mathcal{V}_n([0,1])$ the set
    \[ \mathcal{V}_n([0,1]) = \left\{p \in [0,1]^n \mid \sum_{u = 0}^{n-1} p[u] = 1 \right\}\]
\end{definition}
By definition, we have a bijection
\begin{align*}
    \psi : \mathcal{V}_n([0,1])^2 &\rightarrow \mathcal{M}_n^{\phi^{CM}}([0,1]) \\
    (p^{out}, p^{in}) &\mapsto \mathbb{Q} = p^{out} \cdot (p^{in})^T
\end{align*}
Let's consider a probability distribution $\mathbb{Q} \in \mathcal{M}_n^{\phi^{CM}}([0,1])$, and $p^{out}$, $p^{in} \in \mathcal{V}_n([0,1])^2$ such that $\forall u,v, \mathbb{Q}[u,v] = p^{out}[u] \cdot p^{in}[v]$, then
\begin{align}
    \label{cfm_f}
    \mathrm{f}(\mathbb{Q},E) = &-\sum_{i = 1}^m \log_2(\mathbb{Q}[e_i]) + \sum_{u,v} \mathbb{Q}[u,v]\log_2(\mathbb{Q}[u,v]) \nonumber \\
    = &-\sum_{i = 1}^m \log_2(p^{out}[u_i]\cdot p^{in}[v_i]) + \sum_{u,v} (p^{out}[u] \cdot p^{in}[v])\log_2(p^{out}[u] \cdot p^{in}[v]) \nonumber\\
    = &-\sum_{i = 1}^m \log_2(p^{out}[u_i]) + \sum_u \sum_v p^{out}[u] \cdot p^{in}[v] \cdot \log_2(p^{out}[u]) + \nonumber\\
    &- \sum_{i=1}^m \log_2(p^{in}[v_i]) + \sum_u \sum_v p^{out}[u] \cdot p^{in}[v] \cdot \log_2(p^{in}[v]) \nonumber\\
    = &-\sum_{i = 1}^m \log_2(p^{out}[u_i]) + \sum_u p^{out}[u] \cdot \log_2(p^{out}[u]) + \nonumber\\
    &- \sum_{i=1}^m \log_2(p^{in}[v_i]) + \sum_v p^{in}[v] \cdot \log_2(p^{in}[v])
\end{align}
Hence, if we introduce
\begin{gather*}
    K_u = \#\{k \in \llbracket 1, m \rrbracket, u_k = u\} \\
    K_v = \#\{k \in \llbracket 1, m \rrbracket, v_k = v\}
\end{gather*}
following the same reasoning as in Annex \ref{minimum_proof}, we can define
\begin{align*}
    \mathrm{g}_E^{out} : \mathcal{V}_n([0,1]) &\to \mathbb{R} \\
        p &\mapsto \sum_u (p[u] - K_u) \cdot \log_2(p[u])
\end{align*}
\begin{align*}
    \mathrm{g}_E^{in} : \mathcal{V}_n([0,1]) &\to \mathbb{R} \\
        p &\mapsto \sum_v (p[v] - K_v) \cdot \log_2(p[v])
\end{align*}
They both have a unique minimum which we denote respectively $p^{out*}(E)$ and $p^{in*}(E)$. Then, we define 
\[ \mathbb{Q}_{\phi^{CM}}^*(E) = \psi(p^{out*}(E), p^{in*}(E))\]
Let's show that $\mathbb{Q}_{\phi^{CM}}^*(E)$ is the unique minimum of $\mathrm{f}$ over $\mathcal{M}_n^{\phi^{CM}}([0,1])$. Let $\mathbb{Q} \in \mathcal{M}_n^{\phi^{CM}}([0,1])$ such that $\mathrm{f}(\mathbb{Q}, E) \leq \mathrm{f}(\mathbb{Q}_{\phi^{CM}}^*(E), E)$. Let $p^{out} \in \mathcal{V}_n([0,1])$ and $p^{in} \in \mathcal{V}_n([0,1])$ such that $\mathbb{Q} = \psi(p^{out}, p^{in})$. According to equation \ref{cfm_f}, 
\[ \mathrm{f}(\mathbb{Q}, E) = \mathrm{g}_E^{out}(p^{out}) + \mathrm{g}_E^{in}(p^{in})\]
So, by definition of $\mathbb{Q}$,
\[ \mathrm{g}_E^{out}(p^{out}) + \mathrm{g}_E^{in}(p^{in}) \leq \mathrm{g}_E^{out}(p^{out*}) + \mathrm{g}_E^{in}(p^{in*}) \]
Which implies that $p^{out} = p^{out*}$ and $p^{in} = p^{in*}$, and thus that $\mathbb{Q} = \mathbb{Q}_{\phi^{CM}}^*(E)$. So $\mathbb{Q}_{\phi^{CM}}^*(E)$ is the unique minimum of $\mathrm{f}(\mathbb{Q},E)$ over $\mathcal{M}_n^{\phi^{CM}}([0,1])$.
\subsection{Proof of convergence}
\label{annex:proofcv}
We prove the following result:
\begin{center}
\fbox{\begin{minipage}{0.95\textwidth}
Let $(e_i)_{i \in \mathbb{N}}$ be a sequence of independent and identically distributed random variables following $\mathbb{P}_0 \in \mathcal{M}_n^{\bullet}([0,1])$. 

\[\forall \phi, \mathbb{Q}_{\phi}^*(e_1, \dots, e_x) \underset{x \rightarrow \infty}{\longrightarrow} \underset{\mathbb{Q} \in Prob\_mat_{\phi}}{\mathrm{argmin}} \mathbb{H}(\mathbb{P}_0, \mathbb{Q})\]
\end{minipage}}
\end{center}

Let $\mathbb{P}_0 \in \mathcal{M}_n^{\bullet}([0,1])$.

Let $(e_i)_{i \in \mathbb{N}}$ be a sequence of independent and identically distributed random variables following $\mathbb{P}_0$.

Let's consider the function \[f(\mathbb{Q}, x) = -\sum_{i=1}^x \mathrm{log}_2(\mathbb{Q}[e_i]) + \sum_{u,v} \mathbb{Q}[u,v] \mathrm{log}_2(\mathbb{Q}[u,v])\]

We want to show that: \[\forall \phi, \underset{\mathbb{Q} \in \mathcal{M}_n^{\phi}([0,1])}{\text{argmin}} f(\mathbb{Q},x) \underset{x \rightarrow \infty}{\longrightarrow} \underset{\mathbb{Q} \in \mathcal{M}_n^{\phi}([0,1])}{\text{argmin}} \mathbb{H}(\mathbb{P}_0, \mathbb{Q})\]

Let $\phi$ be an hyperparameter and $\mathbb{Q} \in \mathcal{M}_n^{\phi}([0,1])$. Following the weak law of large numbers
\[-\frac{1}{x}\sum_{i = 1}^{x} \mathrm{log}_2(\mathbb{Q}[e_i]) \underset{x \rightarrow \infty}{\longrightarrow} \mathbb{H}(\mathbb{P}_0,\mathbb{Q})\]
Hence
\[\frac{1}{x} f(\mathbb{Q}, x) - \mathbb{H}(\mathbb{P}_0, \mathbb{Q})\underset{x \rightarrow \infty}{\longrightarrow} 0\]
So if we consider the sequence of functions 
\begin{align*}
    g_x \colon \mathcal{M}_n^{\phi}([0,1]) &\to \mathbb{R} \\
                \mathbb{Q} &\mapsto \frac{1}{x} f(\mathbb{Q}, x) - \mathbb{H}(\mathbb{P}_0, \mathbb{Q})
\end{align*}
it converges point-wise toward 0. As it is an equicontinuous family of functions defined on a compact set of $\mathrm{R}^n$, it converges uniformly toward 0. This means that
\begin{equation}
\label{first_ineq}
\forall \delta > 0, \exists A \in \mathbb{R}^+, \forall \mathbb{Q} \in \mathcal{M}_n^{\phi}([0,1]), \forall x \geq A, \left|\frac{1}{x}f(\mathbb{Q},x) - \mathbb{H}(\mathbb{P}_0, \mathbb{Q}) \right| < \delta
\end{equation}

What is more, if we let $\mathbb{P}' = \underset{\mathbb{Q} \in \mathcal{M}_n^{\phi}([0,1])}{\text{argmin}} \mathbb{H}(\mathbb{P}_0, \mathbb{Q})$. $\mathbb{H}$ is a strictly convex function of $\mathbb{Q}$ so
\begin{equation}
\label{second_ineq}
\forall \epsilon > 0, \exists \delta > 0, \forall \mathbb{Q} \in \mathcal{M}_n^{\phi}([0,1]), |\mathbb{H}(\mathbb{P}_0,  \mathbb{Q}) - \mathbb{H}(\mathbb{P}_0, \mathbb{P}')| < \delta \Rightarrow |\mathbb{Q} - \mathbb{P'}| < \epsilon
\end{equation}

With those two inequalities, we can proceed to the convergence demonstration. Let $\epsilon > 0$, $\delta$ such as in equation \ref{second_ineq}, $A$ such as in equation \ref{first_ineq} with $\frac{\delta}{3}$, and $x \geq A$. Let $\mathbb{Q}(x) = \underset{\mathbb{Q} \in \mathcal{M}_n^{\phi}([0,1])}{\text{argmin}} \frac{1}{x} f(\mathbb{Q},x)$. Because of equation \ref{first_ineq}, we have that
\begin{gather*}
\left| \frac{1}{x}f(\mathbb{Q}(x), x) - \mathbb{H}(\mathbb{P}_0, \mathbb{Q}(x))\right| < \frac{\delta}{3} \\
\left| \frac{1}{x}f(\mathbb{P'}, x) - \mathbb{H}(\mathbb{P}_0, \mathbb{P'})\right| < \frac{\delta}{3}
\end{gather*}

Thus, if $|\mathbb{H}(\mathbb{P}_0, \mathbb{Q}(x)) - \mathbb{H}(\mathbb{P}_0, \mathbb{P'})| \geq \delta$:
\begin{align*}
\frac{1}{x}f(\mathbb{Q}(x), x) &\geq \mathbb{H}(\mathbb{P}_0, \mathbb{Q}(x)) - \frac{\delta}{3} \\
&\geq \mathbb{H}(\mathbb{P}_0, \mathbb{P'}) + \frac{2\delta}{3} \\
&> \mathbb{H}(\mathbb{P}_0, \mathbb{P'}) + \frac{\delta}{3} \\
&> \frac{1}{x} f(\mathbb{P'},x)
\end{align*}
Which contradicts the definition of $\mathbb{Q}(x)$. Thus $|\mathbb{H}(\mathbb{P}_0, \mathbb{Q}(x)) - \mathbb{H}(\mathbb{P}_0, \mathbb{P'})| < \delta$, and because of equation \ref{second_ineq}
\[ |\mathbb{Q}(x) - \mathbb{P'}| < \epsilon \]
Which proves that:
\[\underset{\mathbb{Q} \in \mathcal{M}_n^{\phi}([0,1])}{\text{argmin}} f(\mathbb{Q},x) \underset{x \rightarrow \infty}{\longrightarrow} \underset{\mathbb{Q} \in \mathcal{M}_n^{\phi}([0,1])}{\text{argmin}} \mathbb{H}(\mathbb{P}, \mathbb{Q}) \]
And as this is true for any hyperparameter $\phi$, the result is proved.
\subsection{Description length computation}
\label{annex:dl}
To compute the description length of a sequence $E$, a fundamental result in information theory states that, if a source (let's call her Alice) draws messages independently at random from a set $\Omega$ following a probability distribution $\mathbb{Q}$ and then transmit them to a destination (Bob) over a binary channel, then the code $C_{\mathbb{Q}}: \Omega \rightarrow [0,1]^*$ which minimizes the expected length of the total message $\mathbb{E}_{x \in \Omega}[|C(x)|]$ will be such that:
\[ \forall x \in \Omega, |C(x)| = -\log_2(\mathbb{Q}[x])\]
Therefore, if we suppose that all edges $e_i \in E$ were generated independently following a probability distribution $\mathbb{Q}$, we obtain that:
\begin{align*}
    D(E|\mathbb{Q}) &= -\log_2(\mathbb{Q}[E]) \\
                    &= -\log_2\left(\prod_{k=1}^m \mathbb{Q}[e_k]\right) \\
                    &= -\sum_{i=1}^m \log_2(\mathbb{Q}[e_k]) \\
\end{align*}
\subsection{Bayesian inference}
\label{annex:bay_inf}
We have defined the estimation $\mathbb{Q}_{\phi}^*(E)$ as the model which allows for the best compression of $E$. Yet, if we consider $\mathcal{M}_n^{\phi}([0,1])$ as the set of models which could have been used to generate $E$, $\mathbb{Q}_{\phi}^*(E)$ can also be interpreted as the most likely hypothesis among them. 

According to Bayes' theorem, the probability that a model $\mathbb{Q} \in \mathcal{M}_n^{\phi}([0,1])$ was the one used to generate the edge sequence $E$ is
\begin{equation*}
    \mathbb{P}_{\phi}[\mathbb{Q}|E] = \frac{\mathbb{P}[E|\mathbb{Q}] \times \bar{\mathbb{P}_{\phi}}[\mathbb{Q}]}{\mathbb{P}[E]}
\end{equation*}
Therefore, as $\mathbb{P}[E]$ does not depend on $\mathbb{Q}$, 
\begin{equation}
    \label{bayes}
    \mathbb{Q}_{\phi}^*(E) = \underset{\mathbb{Q} \in \mathcal{M}_n^{\phi}([0,1])}{\mathrm{argmax}} \mathbb{P}[E|\mathbb{Q}] \times \bar{\mathbb{P}_{\phi}}[\mathbb{Q}]
\end{equation}

In practice, it means that if we infer the most likely model for an empty sequence, $\mathbb{Q}_{\phi}^*[\emptyset]$ will be the highest entropy model within $\mathcal{M}_n^{\phi}([0,1])$. On the other hand, as we have more and more observations, the sequence $E$ becomes longer and the influence of the prior distribution $\bar{\mathbb{P}_{\phi}}[\mathbb{Q}]$ becomes negligible. As the probability to generate an edge $(u,v)$ with a model $\mathbb{Q}$ is simply $\mathbb{Q}[u,v]$ and edges are assumed to be independent, this equation becomes
\begin{equation*}
        \mathbb{Q}_{\phi}^*(E) = \underset{\mathbb{Q} \in \mathcal{M}_n^{\phi}([0,1])}{\mathrm{argmax}} \: \prod_{i=1}^m \mathbb{Q}[e_i] \times \frac{1}{Z_{\phi}} \times 2^{\mathbb{S}[\mathbb{Q}]}
\end{equation*}
To perform the maximization, it is simpler to consider the logarithm of this expression. As $\log_2$ is a monotonous function, it does not change the value of $\mathbb{Q}_{\phi}^*(E)$.
\begin{align*}
    \mathbb{Q}_{\phi}^*(E) &= \underset{\mathbb{Q} \in \mathcal{M}_n^{\phi}([0,1])}{\mathrm{argmax}} \: \log_2 \left(\prod_{i=1}^m \mathbb{Q}[e_i] \times \frac{1}{Z_{\phi}} \times 2^{\mathbb{S}[\mathbb{Q}]} \right) \\
    &= \underset{\mathbb{Q} \in \mathcal{M}_n^{\phi}([0,1])}{\mathrm{argmax}} \: \sum_{i=1}^m \log_2(\mathbb{Q}[e_i]) + \mathbb{S}[\mathbb{Q}] \\
    &= \underset{\mathbb{Q} \in \mathcal{M}_n^{\phi}([0,1])}{\mathrm{argmin}} \: - \sum_{i=1}^m \log_2(\mathbb{Q}[e_i]) - \mathbb{S}[\mathbb{Q}]
\end{align*}
\subsection{Edge prediction probability}
\label{edge_prediction_probability}
\begin{figure}
    \centering
    \includegraphics[width=\textwidth]{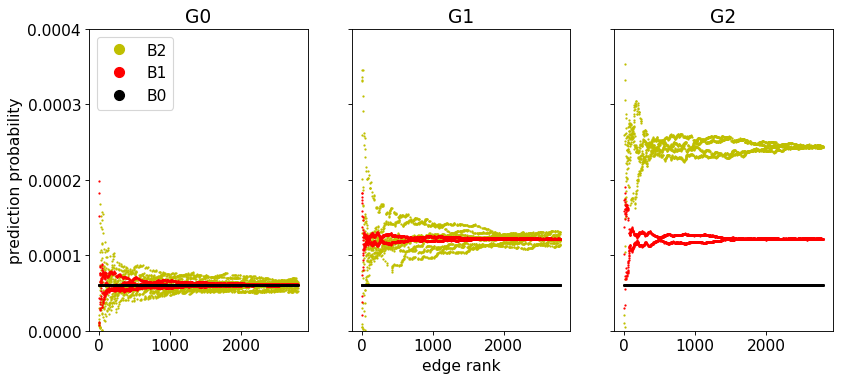}
    \caption{For each graph, we plot the edge prediction probability $\mathbb{Q}^*_{\phi^B}(e_1, \dots, e_{k-1})[e_k]$ against $k$. Black dots corresponds to the model learned with partition $B_0$, red dots with partition $B_1$ and yellow dots with partition $B_2$. As the number of observed edges grows, the prediction converges to a value which depends on $G$ and $B$. When the learning partition is coarser than the original partition, the prediction probability converges to a lower value. When it is finer, it converges toward the same value, but more slowly.}
    \label{sbm_1}
\end{figure}

Before looking at the description length, we investigate how the prediction probability of the next edge evolves as Alice draws more and more edges. We consider three stochastic blockmodels on $n = 128$ nodes based on three partitions $B_0$, $B_1$ and $B_2$. $B_0$ is made of a single block of size $128$, $B_1$ of two blocks of size $64$ and $B_2$ of four blocks of size $32$, obtained by dividing $B_1$'s block in half. The three SBMs are fully described in Table \ref{sbm_0}. For each SBM, we randomly sample $m = 2800$ edges and thus obtain three graphs: $G_0$, $G_1$ and $G_2$. We want to study the edge prediction probability evolution depending on the constraints used to learn the model. Thus, for each of the three graphs, and each of the three hyperparameters $\phi^{B_0}$, $\phi^{B_1}$, $\phi^{B_2}$, we plot the evolution of the prediction probability $\mathbb{Q}^*_{\phi}(e_1, \dots, e_{k-1})[e_k]$ against $k$ in Figure~\ref{sbm_1}.

\begin{table}[t]
    \centering
    \begin{tabular}{c|m{4cm}|l}
        Model & Partition & Block probability matrix \\
        \hline
        & & \\ [0.2ex]
         $S_0$ & $B_0 = \llbracket 1, 128 \rrbracket$ & $M_0 = \frac{1}{n^2} \cdot \begin{bmatrix} 1 \end{bmatrix}$ \\[2ex]
         $S_1$ & $B_1 = \llbracket 1, 64 \rrbracket$, $\llbracket 65, 128 \rrbracket$ & $M_1 = \frac{1}{n^2} \cdot \begin{bmatrix} 2 & 0 \\ 0 & 2 \end{bmatrix}$ \\[2ex]
         $S_2$ & $B_2 = \llbracket 1, 32 \rrbracket$, $\llbracket 33, 64 \rrbracket$, $\llbracket 65, 96 \rrbracket$, $\llbracket 97, 128 \rrbracket$ & $M_2 = \frac{1}{n^2} \cdot \begin{bmatrix} 4 & 0 & 0 & 0 \\ 0 & 4 & 0 & 0 \\ 0 & 0 & 4 & 0 \\ 0 & 0 & 0 & 4 \end{bmatrix}$ \\
    \end{tabular}
    \caption{Three stochastic blockmodels defined as edge probability distributions. }
    \label{sbm_0}
\end{table}

This simple example shows how the level of constraints imposed by the hyperparameter acts on the probability prediction of the next edge. For all three graphs, whatever $k$, the prediction probability based on the null partition $B_0$ is constant at $0.00006$ (black dots). This is logical, as the only probability matrix in $\mathcal{M}^{\phi^{B_0}}_n([0,1])$ is the uniform distribution. Therefore, 
\[\forall k, \mathbb{Q}^*_{\phi^{B_0}}(e_1,\dots, e_{k-1})[e_k] = \frac{1}{n^2} = \frac{1}{128^2} \approx 0.00006 \]
For other hyperparameters (red and yellow dots), the results depend on the graph. On $G_0$, generated with $B_0$ and thus presenting no block structure, models based on more refined partitions do not lead on average to better prediction probabilities than the one based on $B_0$. For some edges their prediction probability is better, but as often it is worse. On average, they have the same prediction power, but the convergence toward the generative probability distribution is slowed down by random fluctuations due to the additional degree of freedom allowed. 

On the other hand, for $G_1$, generated with $B_1$ (two blocks), we observe that refining the partition from one block to two allows the prediction probability to increase quickly. While it remains $\frac{1}{n^2}$ for the hyperparameter $\phi^{B_0}$, it converges to $\frac{2}{n^2}$ for the hyperparameter $\phi^{B_1}$ (red dots). Yet, refining even more the partition is worthless, as illustrated by the $B_2$ partition (yellow dots), with $4$ blocks, which does not bring any improvement on average. Finally, considering $G_2$, we observe that refining the partition brings more and more improvement to the prediction probability. With $B_0$ it remains stable at $\frac{1}{n^2}$, with $B_1$ it rises up to $\frac{2}{n^2}$, and with $B_2$ up to $\frac{4}{n^2}$. This shows that increasing the number of degrees of freedom of the model (\textit{i.e.} reducing the number of constraints of the hyperparameter) is a double-edged sword. As long as it allows the model to better fit correlations that are present in the observations, it leads to better prediction performance. Yet, this comes at the price of a slower convergence of the model. It is the combination of those two effects which allows us to detect both overfitting and underfitting models.
\end{document}